\documentclass[preprint,showpacs,preprintnumbers,amsmath,amssymb]{revtex4}

\usepackage{graphicx}
\usepackage{dcolumn}
\usepackage{bm}

\newcommand{\be}{\begin{eqnarray}}
\newcommand{\ee}{\end{eqnarray}}
\newcommand{\bdm}{\begin{displaymath}}
\newcommand{\edm}{\end{displaymath}}
\newcommand{\nn}{\nonumber}

\begin{document}

\preprint{APS/123-QED}

\title{Minimizing the kinematical effects on LISA's performance}

\author{Ioannis Deligiannis}
\author{Theocharis A. Apostolatos}%
\affiliation{Section of Astronomy, Astrophysics, and Mechanics\\
University of Athens\\ Panepistimiopolis, Zografos, GR-15783,
Athens, Greece }%

\date{\today}

\begin{abstract}

Proper tuning of the orbital characteristics of the three spacecrafts that constitute
the usual triangular configuration of the space-borne gravitational-wave detector LISA, 
could minimize the breathing mode of 
its arm-lengths. Since the three spacecrafts form three pairs of
interferometric arms, we have the freedom to minimize whichever
combination of arm-length variations that might be useful in signal analysis. Thus for any kind of time delay
interferometry (TDI), that is chosen to be used in analysing the data, 
the optimal orbital characteristics could be chosen accordingly, so
as to enhance the performance of the gravitational wave detector.
\end{abstract}

\maketitle

\section{\label{sec:1}
Introduction}

The joint ESA-NASA future mission to launch a spaceborne gravitational wave
antenna, known as LISA (Laser Interferometer Space Antenna), is expected to offer us invaluable information for
the Universe. The detector will be able to monitor low frequency gravitational
waves, the source of which could be either supermassive black hole binaries 
at cosmological distances \cite{Schu86,CutlVali07}, white dwarf and/or neutron star binaries \cite{Ferretal}, 
or primordial waves of cosmological origin \cite{LISA,ChonEfsta}.
Detection of such gravitational waves could be achieved by placing three spacecrafts
into three distinct Earth-like orbits, so that they form an equilateral triangle of 
almost constant size \cite{DhurNayaKoshVine}, and monitoring interferometrically
the tiny variations of distances between any pair of such spacecrafts that are induced by gravitational
waves passing by LISA \cite{LISA1}. Since
the frequency band that this detector is sensitive at is in the region $10^{-4}$ to $10^{-1}$ Hz,
any possible relative motions of spacecrafts that occur at much lower frequencies of order $\sim 10^{-7}$
Hz (due to orbital periodicities) could be filtered out from the signal \cite{CornRubo,02TintEstaArms}.
However, despite the fact that the large beat modes due to  Doppler shifts caused by arm-length variations could
be effectively eliminated, the same arm-length variations could cause unpleasant complications while attempting 
to reduce internal noise by various schemes of time delay interferometry (known as TDI's) \cite{TDI,ValiTDI}. 
This happens because, while the
laser beams travel back and forth across the arms, they pass through each spacecraft at different times due to
arm-length changes; thus any reference frequency variations of local lasers do not exactly cancel 
out in the corresponding combination of signals,
as it would happen if there was no flexing of the  arms. Consequently,
if the variability of the size of the triangular formation could be minimized by suitable adjustment of their
orbital characteristics, this would lead to a suppression of the internal noise of LISA, which, in its turn, would be highly 
beneficial for its performance as a gravitational wave detector.

We start by writing down the position of each spacecraft, accurate up to second order with respect
to its eccentricity. Each spacecraft moves on a slightly elliptical orbit with semi-major axis equal to
1 AU, the plane of which is slightly inclined with respect to the plane of the ecliptic. Finally the orbit
of each spacecraft is rotated by an angle of $\pm 2 \pi/3$, with respect to the other two orbits, on the ecliptic plane.
The initial position of the three spacecrafts on their corresponding orbits is such
that they form an equilateral triangle that remains equilateral to first order with respect to the eccentricity $e$ of
the orbits. The proposed configuration of LISA, that is widely used in corresponding analyses \cite{DhurNayaKoshVine}, 
assumes that the plane of
the triangular configuration forms an exact $60^\circ$ angle with respect to the ecliptic plane, 
with the triangle being exactly equilateral
initially. Indeed this configuration ensures stable distances between spacecrafts to first order with respect
to the eccentricity of the orbits $e$. However, this configuration leads to a breathing mode of the arms
with amplitude of order $e^2$ \cite{NayaKoshDhurVine}. 
This arm flexing end up generating noise, by one way or another, in detector's output
\cite{DhurNayaKoshVine,CornHell,ValiTDI}.

It is to verify that each spacecraft has 3 extra degrees of freedom that could be
used to minimize whichever arm-length variation might one choose. 
Of course this fine tuning of the orbital characteristics
should be one order of $e$ higher than the initially proposed value of the characteristics themselves, 
so that the invariance of the arm-lengths to order $e$ is not
destroyed. These nine, altogether, degrees of freedoms could be chosen to be the eccentricities of the orbits, 
their inclinations,
and the initial angle position of each spacecraft along 
its own slightly eccentric and inclined orbit. The choice could be
such that the initial configuration deviates from being equilateral to order $e$, and/or its plane inclination deviates from
the $60^\circ$ to order $e$, as well. On the other hand such an adjustment could reduce
the time-variation of any arm-length, or the relative variation of any pair of arm-lengths, at its minimum value.

The rest of the paper is organized as follows. In Sec. \ref{sec:2} we repeat the basic calculations, in a form that
could be later extended to higher order with respect to $e$, that
show why the initially designed configuration leads to a time-independent equilateral triangular configuration
to order $e$. By expanding the spacecraft distances to order $e^2$, we compute the time variation of the arm-lengths at
this order. In Sec. \ref{sec:3} we introduce nine extra parameters of order $e^2$ that modify the initial
positions of the three spacecrafts. Then we calculate once again the distances between each pair of spacecrafts
as a function of time, which are now parametrized by the six (out of nine) essential parameters that determine the initial location of
the spacecrafts. Finally, in Sec. \ref{sec:4} we show how we could optimally choose the fine tuning parameters
so as to get the minimum contribution to noise from the variations of the arm-lengths, depending on the TDI
scheme that one might choose to use in signal extraction.

\section{\label{sec:2}The initially designed orbits}

\begin{figure}
\begin{center}
\includegraphics[width=.8\textwidth]{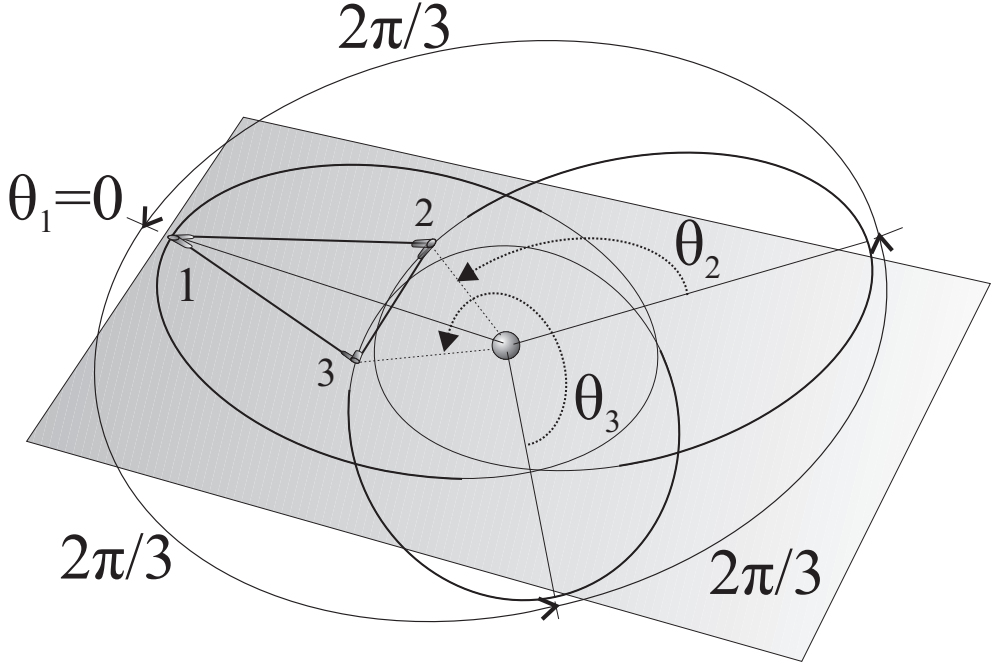}
\caption{The figure depicts the orbits of the three spacecrafts labeled 1, 2, and 3 respectively. The triangular
configuration is shown at an instance when spacecraft 1 is at its aphelion ($\theta_1=0$)
while the other spacecrafts are located at angles $\theta_2 = 2 \pi/3+{\cal O}(e)$, and
$\theta_3 = 4 \pi/3+{\cal O}(e)$, respectively. The three elliptical orbits are inclined with respect to
the ecliptic plane (shown in the figure); thus the part of the ellipses that lie above the ecliptic
plane are drawn thicker than the parts that lie below. Also the three orbits are rotated with respect
to each other by exactly $\pm 2 \pi/3$, as shown in the figure. The inclinations and the eccentricities
presented in this figure are highly exaggerated to clearly illustrate the geometry of the orbits.} \protect\label{fig:1}
\end{center}
\end{figure}

\begin{figure}
\begin{center}
\includegraphics[width=.8\textwidth]{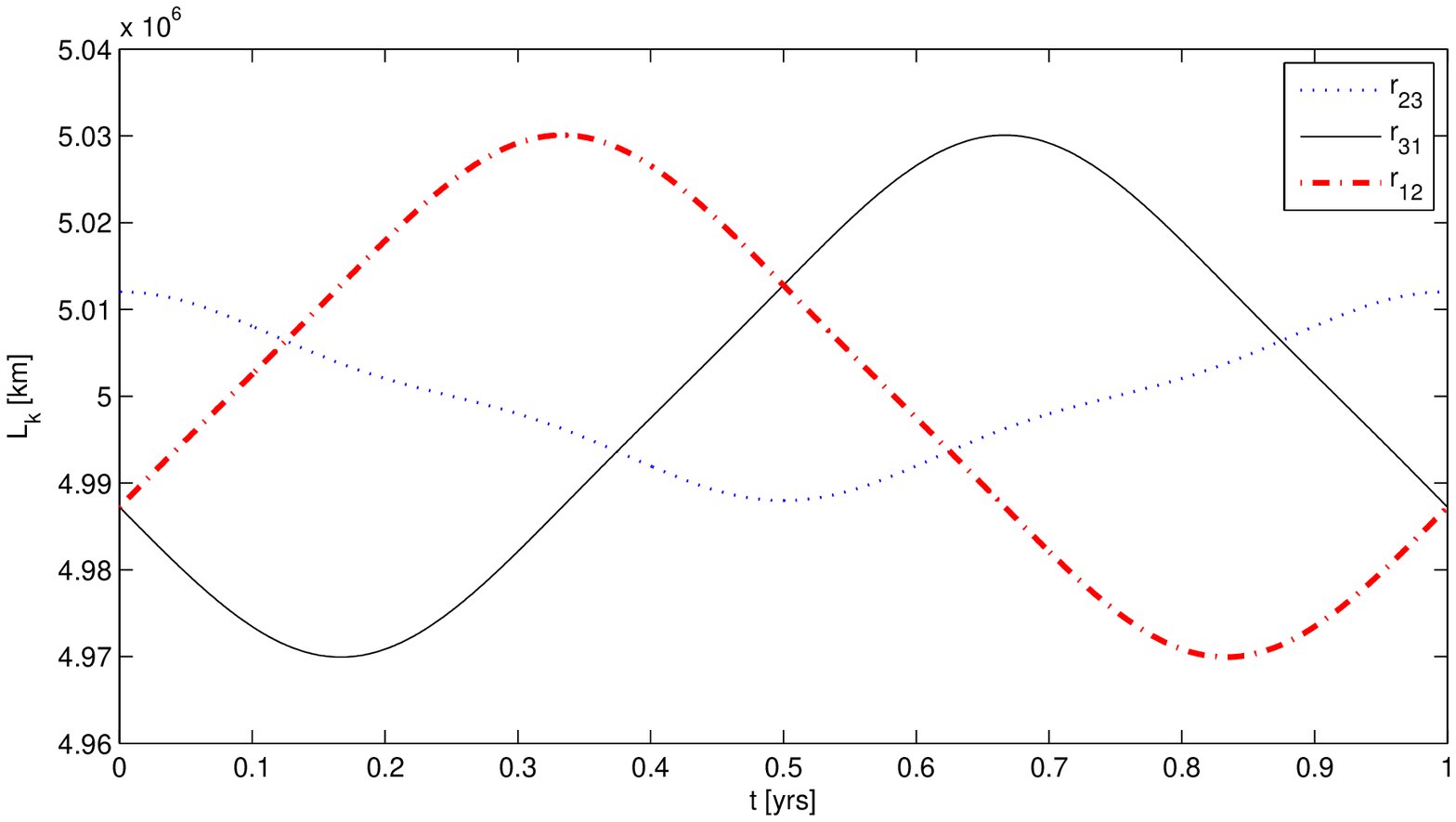}
\caption{This diagram shows the variation of the three arm-lengths
for the configuration of LISA that was initially designed.
The range of arm-lengths' variation is $6.6 \times 10^4$ km, for $r_{13}$, and $r_{12}$,
and $2.1 \times 10^4$ km, for $r_{23}$, respectively. The first two pairs have similar time evolution due 
to symmetric arrangements of the corresponding sides of the equilateral triangle. Actually $r_{12}(t)=r_{13}(-t)$,
if $t=0$ corresponds to configuration shown in Figure \protect\ref{fig:2}. } \label{fig:2}
\end{center}
\end{figure}

The orbit of each spacecraft is a Keplerian ellipse, if one ignores the gravitational attraction due to
planets (mainly Earth, and Jupiter) and relativistic effects. Assuming that all spacecrafts are moving
on orbits that have semi-major axis of 1 AU, so that all return at their initial positions after one year, and
are not drifting away secularly with respect to Earth, their distance from the Sun could be written as
\be
r=a(1+e \cos\xi)
\ee
where $a$, the semi-major axis, is common for all spacecrafts and is equal to 1 AU. The eccentricity, $e$,
is common for all spacecrafts, as well, for the proposed configuration and is the only parameter that
uniquely characterizes all three orbits (see below). Finally, $\xi$ is the so called eccentric anomaly, which is an angle
parameter that determines the
position of the spacecraft along its elliptical orbit. Now the orbital plane of each spacecraft
is inclined with respect to the ecliptic plane by an angle $\lambda e$, where $\lambda$ is a number
that will be determined by demanding the configuration of the three spacecrafts to be an 
equilateral triangle with invariant size
to first order with respect to $e$. The three inclined orbital planes are rotated by an angle
of $\pm 2 \pi/3$, with respect to each other, on the ecliptic plane (see Fig.~\ref{fig:1}).
By using the eccentric anomaly to describe the position of each spacecraft along its orbit, instead of the polar angle
$\theta$, the calculations that lead to the distance between two spacecrafts as a function of time are made easier
since by conservation of angular momentum the angular position $\theta$ on a keplerian orbit 
as a function of time is given by
\be
\label{thetat}
t=\int_{\theta_0}^{\theta(t)} \frac{d \theta}{\dot \theta}=
\frac{m a^2 (1-e^2)^2}{L} \int_{\theta_0}^{\theta(t)} \frac{d \theta}{(1-e \cos\theta)^2},
\ee
while by using the eccentric anomaly parameter $\xi$ which is related to $\theta$ via
\be
\label{xitheta}
1+e \cos \xi \equiv \frac{1-e^2}{1-e\cos\theta},
\ee
the cartesian coordinates of the orbital position are
\be
\begin{split}
x&=&r \cos \theta=a(\cos\xi+e), \\
y&=&r\sin\theta=a \sqrt{1-e^2} \sin\xi,
\label{xycoords}
\end{split}
\ee
and the integral of Eq.~(\ref{thetat}) is easily computed to yield
\be
\xi-\xi_0+e(\sin\xi-\sin\xi_0)= \frac{L}{m a^2 \sqrt{1-e^2}} t=\omega t,
\label{xi(t)}
\ee
(c.f.~\cite{Landau}).
The last expression is easier to use, than Eq.~(\ref{thetat}), to compute $t(\xi)$, in contrast to $t(\theta)$.
The above transcendental equation could not be exactly inverted in a closed analytical form. However,
this could be achieved in the form of a power expansion
with respect to $e$, yielding
\be
\xi=\omega t+\xi_0 &+& \left( \sin(\omega t+\xi_0)-\sin\xi_0 \right)\times\nn \\
&&\left(-e+e^2 \cos(\omega t+\xi_0)+O(e^3)\right).
\ee
The cartesian coordinates of each spacecraft on the heliocentric system, with the $z$-axis perpendicular to the ecliptic plane, 
are thus given by
\be
{\bf x}^{(i)}= R^{(i)}_{\textrm{rot}} R^{(i)}_{\textrm{inc}} {\bf x}^{(i)}_{\textrm{op}}
\ee
where ${\bf x}^{(i)}_{\textrm{op}}$ is the position of the $i$-th spacecraft on its orbital plane, that is
\be
{\bf x}^{(i)}_{\textrm{op}}=\left(
\begin{array}{c}
r^{(i)} \cos \theta^{(i)} \\
r^{(i)} \sin \theta^{(i)} \\
0
\end{array} \right)=
\left(
\begin{array}{c}
a(e + \cos \xi^{(i)}) \\
a \sqrt{1-e^2} \sin \xi^{(i)} \\
0
\end{array} \right)
\ee
where the expressions in the last array come from Eqs.~(\ref{xycoords}).
The matrix $R^{(i)}_{\textrm{inc}}$ produces an inclination of the orbital plane
by an angle $\lambda e$ with respect to ecliptic plane around $y$-axis. Therefore,
\be
R^{(i)}_{\textrm{inc}}=\left(
\begin{array}{ccc}
\cos(\lambda e) & 0 & \sin(\lambda e)\\
0               & 1 & 0\\
-\sin(\lambda e)& 0 & \cos(\lambda e)\\
\end{array} \right).
\label{Rinc}
\ee
Finally, $R^{(i)}_{\textrm{rot}}$ is the
matrix that rotates the orbit of the $i$-th spacecraft by $\Phi_1=0, \Phi_2=2 \pi/3$, and
$\Phi_3=4 \pi/3$, respectively, on the ecliptic plane, that is
\be
R^{(i)}_{\textrm{rot}}=\left(
\begin{array}{ccc}
\cos\Phi_i  & \sin\Phi_i & 0\\
-\sin\Phi_i & \cos\Phi_i & 0\\
0           & 0          & 1
\end{array} \right).
\label{Rrot}
\ee
By combining the coordinates of each spacecraft given above, and choosing the initial angular
positions $\xi^{(i)}_0$ (the superscript $^{(i)}$ refers to the $i$-th spacecraft) to be
\be
(\xi^{(1)}_0,\xi^{(2)}_0,\xi^{(3)}_0)=(0,\frac{2 \pi}{3}- \frac{\sqrt{3}}{2} e,\frac{4 \pi}{3}+\frac{\sqrt{3}}{2} e),
\ee
along with the common inclination parameter $\lambda=\sqrt{3}$,
we obtain the following time depending distance of each pair of spacecrafts to
second order with respect to $e$:
\begin{widetext}
\be
r_{12}(t)&=&2 \sqrt{3}a e- \frac{\sqrt{3}}{32}\left(
19 \cos(\omega t)
-19\sqrt{3} \sin(\omega t)
-2 \cos(3\omega t) \right) a e^2 + O(e^3), \\
r_{23}(t)&=&2 \sqrt{3} a e+ \frac{\sqrt{3}}{16} \left(
7 \cos(\omega t)
+ \cos(3\omega t) \right) a e^2 + O(e^3), \\
r_{31}(t)&=&2 \sqrt{3} a e- \frac{\sqrt{3}}{32}\left(
19 \cos(\omega t)
+19\sqrt{3} \sin(\omega t)
-2 \cos(3\omega t) \right) a e^2 + O(e^3) .
\ee
\end{widetext}
The choice of initial parameters for the three spacecrafts, mentioned above, is the one that leads to the essential
advantage of the configuration; that is to keep the distances equal, and time-invariant to
lowest order (first order) with respect to parameter $e$. However,
the arm-lengths vary with time to order $e^2$. The magnitude of the arm-length oscillation is thus
of order $10^4$ km as shown in Figure \ref{fig:2}.

\section{\label{sec:3} Lowering the amplitude of arm flexing}

In this section we will use our freedom of choosing the initial positioning
of the three spacecrafts, in order to minimize whichever distance variation we might
like. Of course by placing two spacecrafts
on a circular orbit around the Sun the distance between them will remain fixed,
but in order to sense the quadrupole nature of a gravitational wave we need
al least one more spacecraft that is not along the same line of the former two ones.
Thus, by placing a third spacecraft in an orbit that is inclined with respect to the orbit the other pair,
we will have a time varying arm-length between the third and each one of the other two
spacecrafts, the overall variation of which will be of the order of the initial arm-length.
The clever symmetric configuration of the three slightly inclined and
slightly non-circular orbits that was discussed in the previous section manages to
keep all the arm-lengths constant, at least at the order of magnitude
of the arm-lengths themselves. On the other hand by trying to achieve
a very symmetric configuration we have ignored any possible freedom
we still have to shift our orbits so as to reduce the variation of
distances to even higher order.

Speaking of freedom of initial positioning, 
the three elliptical orbits could be a little different, with respect to eccentricity
and with respect to inclination, with each other. Of course the fine
tuning of the corresponding six parameters will be of order $e^2$, so as to keep the
main characteristic of the configuration; namely the constant value of the
arm-lengths at order $e$. Furthermore one could also loosen the
exact symmetric placement of the orbits by a rotating angle of $\pm 2 \pi/3$ with respect to each other.
However these angles could be kept invariant, and we could alternatively  adjust the initial
angular positions $\xi^{(i)}_0$ to order $e^2$.

Henceforth we will assume that
the three orbits are characterized by the following orbital parameters:
\be
\begin{array}{ccc}
e_1=e ,& e_2=e+\alpha_2 e^2 ,& e_3= e+\alpha_3 e^2, \\
\lambda_1= \sqrt{3}+\beta_1 e ,&
\lambda_2= \sqrt{3}+\beta_2 e ,&
\lambda_3= \sqrt{3}+\beta_3 e.
\end{array}
\ee
We have actually chosen the eccentricity of the first spacecraft's orbit as
a reference for the other two. This eccentricity will play the role of the small expansion parameter
in all positional expressions. Next we define the initial positions of
the spacecrafts along these slightly deformed, and differently inclined orbits
by the initial angles
\be
\begin{split}
&\xi^{(1)}_0=0,\\
&\xi^{(2)}_0=\frac{2 \pi}{3}-\frac{\sqrt{3}}{2}e+\gamma_2 e^2,\\
&\xi^{(3)}_0=\frac{4 \pi}{3}+\frac{\sqrt{3}}{2}e+\gamma_3 e^2,
\end{split}
\ee
where once again by setting $\xi^{(1)}_0$ equal to zero, we have just adjusted
the initial time of the configuration to coincide with the aphelion of spacecraft 1. 
As we see now there are only 7 parameters
that we could tune to make the configuration comply with our demands. The assumed shifts of the
orbital parameters of the three spacecrafts are independent to each other, and thus they
exhaust the whole freedom we have to shift the initial positions of the spacecrafts.

Actually the proposed configuration that forms initially  an exact equilateral triangle inclined by $60^\circ$
with respect to the equatorial orbit corresponds to some specific relations between these 7 new parameters since, as is
shown in Figure~\ref{fig:2}, omission of all these parameters does 
not lead to three equal arm-lengths at $t=0$.

\begin{figure}
\begin{center}
\includegraphics[width=1\textwidth]{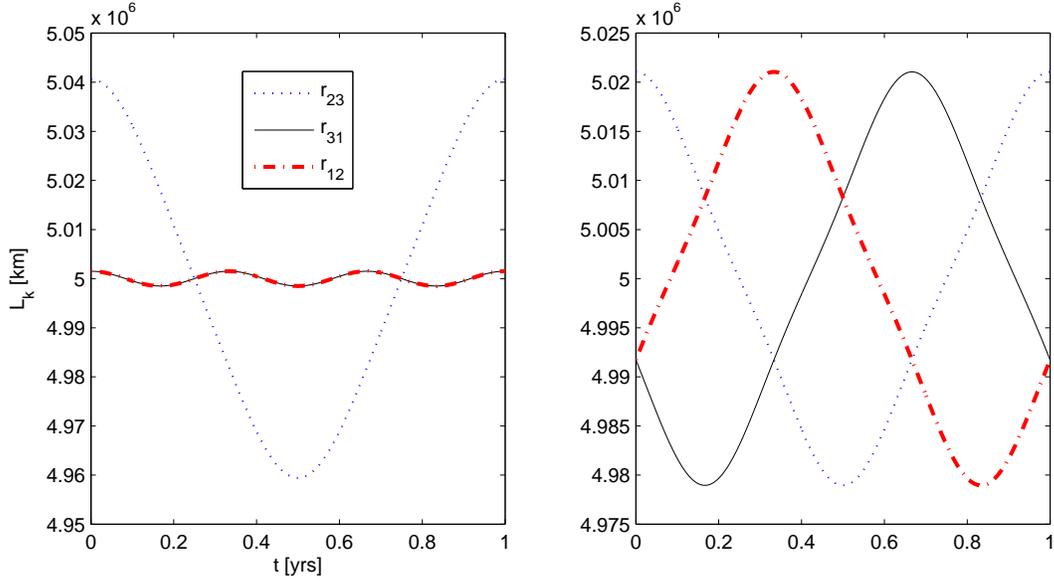}
\caption{ This diagram shows the time varying length of the three arms when
the orbital characteristics of the three spacecrafts are a little bit
different than the ones of the initial design. The left diagram
shows the extreme case $\chi_2=\chi_3=0$, where two of the arms are
oscillating at a small fraction of the oscillation amplitude of the initial configuration (only
by $0.2\times10^4$ km in amplitude), while the third arm is oscillating with an amplitude
which is more than 3 times larger than the amplitude in the initial configuration.
The right diagram shows a more balanced fine tuning of the parameters
($\chi_2=\chi_3=39$) where all arms are oscillating
with an amplitude $\sim 60 \%$ lower
than the maximum amplitudes of initial configuration.} \protect\label{fig:3}
\end{center}
\end{figure}

Repeating once again the computations of the previous section with all seven new
parameters introduced in the formulae for ${\bf x}_{\textrm{op}}^{(i)}$, $R_{\textrm{inc}}^{(i)}$,
and $\xi^{(i)}_0$, we obtain the following expressions for the three distances
as functions of time:
\be
\left(\begin{array}{c}
r_{12}(t)\\r_{31}(t) \\r_{23}(t)
\end{array}\right)=
2 \sqrt{3} a e \left(\begin{array}{c}
1\\1 \\1
\end{array}\right)+ \frac{ a e^2 }{32\sqrt{3}} {\bf A}(t) + \textrm{O}(e^3)
\ee
with
\be
{\bf A}(t)=
{\bf C}_0+
{\bf C}_1 \cos(\omega t)+
{\bf C}_2 \cos(2 \omega t)+
{\bf C}_3 \cos(3 \omega t)+
{\bf S}_1 \sin( \omega t)+
{\bf S}_2 \sin(2 \omega t) .
\ee
The $3 \times 1$ column matrices ${\bf C}_k$'s and ${\bf S}_k$'s, through which we have decomposed
the time-depending second-order column matrix ${\bf A}(t)$,
correspond to the following expressions that depend only on the seven
fine-tuning parameters introduced above:
\be
\label{thecomponents}
\begin{array}{cc}
{\bf C}_0=12
\left(\begin{array}{c}
8 \alpha_2+\sqrt{3}(\beta_1+\beta_2)\\
8 \alpha_3+\sqrt{3}(\beta_1+\beta_3)\\
8 (\alpha_2+\alpha_3)+\sqrt{3}(\beta_2+\beta_3)
\end{array}\right) ,
{\bf C}_1=
\left(\begin{array}{c}
-57-8(3 \alpha_2+2\sqrt{3}\gamma_2)\\
-57-8(3 \alpha_3-2\sqrt{3}\gamma_3)\\
42-16(3 (\alpha_2+\alpha_3)+2\sqrt{3}(\gamma_2-\gamma_3))
\end{array}\right), \\
{\bf C}_2=12\sqrt{3}
\left(\begin{array}{c}
\beta_1\\ \beta_1\\ -(\beta_2+\beta_3)
\end{array}\right) ,
{\bf C}_3=6
\left(\begin{array}{c}
1\\ 1\\ 1
\end{array}\right), \\
{\bf S}_1=3
\left(\begin{array}{c}
19 \sqrt{3} + 8 \sqrt{3} \alpha_2 + 16 \gamma_2 \\
-19 \sqrt{3} - 8 \sqrt{3} \alpha_3 + 16 \gamma_3 \\
0
\end{array}\right) ,
{\bf S}_2=12
\left(\begin{array}{c}
\beta_1+2\beta_2 \\
-\beta_1-2\beta_3 \\
\beta_2-\beta_3
\end{array}\right).
\end{array}
\ee
An obvious optimization choice of parameters is $\beta_1=\beta_2=\beta_3=0$,
since then the component of the variation of arms that oscillates
at frequency $2 \omega$ vanishes (${\bf C}_2={\bf S}_2={\bf 0}$)
without affecting the rest time-depending components ${\bf C}_1,{\bf C}_3,{\bf S}_1$. 
On the other hand the only component
of arm-length oscillation with frequency $3 \omega$ cannot be adjusted through
suitable choice of the parameters. Finally the magnitude of the
components corresponding to frequency $\omega$ could be adjusted by varying the value of
specific combinations of $\alpha_2,\gamma_2$, and $\alpha_3,\gamma_3$.
More specifically by using the following replacements
\be
\begin{split}
\chi_2 &= 57+8(3 \alpha_2+2 \sqrt{3} \gamma_2), \\
\chi_3 &= 57+8(3 \alpha_3-2 \sqrt{3} \gamma_3),
\end{split}
\ee
where the two new parameters $\chi_2,\chi_3$ could be adjusted independently
to each other, the lowest-order arm-length variations oscillating at frequency $\omega$
turn out to be
\be
&&\frac{a e^2}{32 \sqrt{3}} \left(
\left(\begin{array}{c}
-\chi_2 \\ -\chi_3 \\ 156-(\chi_2+\chi_3)
\end{array}\right)
\cos(\omega t) + \right.
\left.\sqrt{3}
\left(\begin{array}{c}
\chi_2 \\ -\chi_3 \\ 0
\end{array}\right)
\sin(\omega t) \right)=\\
&&\frac{a e^2}{32 \sqrt{3}}
\left(\begin{array}{c}
2 \chi_2 \cos(\omega t-2 \pi/3) \\
2 \chi_3 \cos(\omega t-4 \pi/3) \\
\left[156-(\chi_2+\chi_3)\right] \cos(\omega t)
\end{array}\right) .
\label{timedepflex1}
\ee
Therefore, all possible optimizations could be done by suitable adjustments of
$\chi_2,\chi_3$ alone, that is, by simultaneous adjustments of $\alpha_i$'s
and $\gamma_i$'s. For example, by choosing $\chi_2=\chi_3=0$, two out
of three arms are oscillating with the lowest possible amplitude, which is a mere
$5 \%$ of the amplitude of the  initially designed configuration. The price
though is that then the third arm is oscillating with an amplitude that is $\sim 3.4$
times higher than the corresponding amplitude of the initial configuration. On the other hand if
we need to use all arms in our signal analysis, we could make a
compromise by a suitable choice of $\chi_i$'s and manage to decrease the
oscillation amplitude of all arms at an optimized ratio. In Figure~\protect\ref{fig:3},
we have plotted the time-varying arm-lengths in a period of one year, for
two choices of $\chi_i$'s.

\section{Reducing the noise in various TDI schemes}
\label{sec:4}

The optimal geometric configuration, that we could achieve by suitable initial positioning of the three spacecrafts, depends
on the specific TDI (time delay interferometry) scheme that we decide to use in order to extract the signal from the
internal noise of local lasers in each spacecraft. For example,
if we had chosen to synthesize the phase-differences in an equal-arm four-link Michelson
scheme interferometry (see \cite{TDI,ValiTDI}) the noise induced by
the breathing mode of the arms would be
\be
2 {\dot C}_1(t) (L_2(t)-L_3(t)),
\ee
assuming that the spacecraft 1 is the corner station of the corresponding Michelson interferometer.
$L_i(t)$ is the time depending length of the $i$-th arm, while 
$C_i(t)$ describes the time-dependent fluctuations of the $i$-th laser frequency (assuming there is only one reference
laser in each spacecraft).
The lasers of LISA mission are designed to have single-sided spectral density of order 
$30 ~\textrm{Hz}/\sqrt{\textrm{Hz}}$.
The initial configuration has a time varying arm-length
difference which has an amplitude of $3.3 \times 10^4 ~\textrm{km}$ (c.f.~Figure~\ref{fig:2}). On the
other hand in the optimized configuration with $\chi_2=\chi_3=0$
although these two arms actually breathe, they have continuously exactly the same length (at least to order $e^2$
which corresponds to our approximations).
The same holds good also for an unequal-arm eight-link Michelson combination,
denoted X by Armstrong, Estabrook, and Tinto \cite{ArmsEstaTint99},
which is a second-generation TDI scheme \cite{ValiTDI}, since the
antisymmetric combination $\dot{L}_2 L_3-\dot{L}_3 L_2$ is again continuously zero (to the same order of approximation),
and thus the internal laser noises cancel out (c.f.~Eq.~(12) of
\cite{ValiTDI}).

Let us examine one more second-generation TDI scheme that is, now, not symmetric with respect to a specific
pair of arms. For example, the eight-link Relay scheme, denoted U in \cite{ArmsEstaTint99},
leads to the following TDI noise due to laser internal noise:
\be
{\dot C}_3(t) [({\dot L}_{1'}+{\dot L}_{1})(L_{3'}+L_{2'})-
               ({\dot L}_{3'}+{\dot L}_{2'})(L_{1'}+L_{1})+
               {\dot L}_{1}L_{1'}-{\dot L}_{1'}L_{1}].
\ee
By expressing the three arm-lengths as $L_i=L_{i'}=L_0+l_i(t)$ where
$L_0$ is of order $a e$ while $l_i(t)$ is of order $a e^2$ and the numbering is such that
it corresponds to the $i$-th component of the column matrix of Eq.~(\ref{timedepflex1})
the above expression for the noise yields
\be
{\dot C}_3(t) \frac{2 L_0 a e^2 \omega}{32 \sqrt{3}}
[(156+\chi_2-2\chi_3)\sin\omega t+ 
\sqrt{3}(2 \chi_2+\chi_3)\cos\omega t]+\textrm{O}(e^4).
\label{Uoptimize}
\ee
To compute the above expression, only the Fourier components corresponding to frequency $\omega$ has been written down, since the
Fourier component of arm breathing with frequency $3 \omega$ is the same for all arms (c.f.~${\bf C}_3$ of Eq.~(\ref{thecomponents})),
and the corresponding terms cancel out. From expression (\ref{Uoptimize}) it is easy to verify that by
choosing $\chi_2=-156/5$ and $\chi_3=312/5$ we could nullify the noise of this TDI scheme to this order; namely $a^2 e^3$.
It should be noted that the specific choice of numbering of the arms with respect to the assumed distance $r_{ij}$ that we
have used in our analysis leads to the specific optimizing parameter values $\chi_{2,3}$ that we have found.

Actually, all second generation eight-link TDI schemes lead to similar expressions for the noise,
which could be written as 
antisymmetric products of arm-length variations and arm-lengths \cite{ValiTDI}. These products could be expressed as a combination of
$\sin\omega t$ and $\cos\omega t$ terms with corresponding factors that depend on the two parameters $\chi_{2,3}$. Therefore
there is always a suitable combination of the $\chi_{2,3}$ parameters that eliminates the noise to that order,
which means that by suitable fine initial positioning of the three spacecrafts we could depress the laser noise
at the level of ${\dot C} a^2 e^4 \omega$. This is the best optimization we could achieve for a specific second generation
eight-link TDI scheme based on the kinematics of LISA.
As an order of magnitude, this means a reduction in the noise of LISA by $e \simeq 1/100$ with respect to a
non-optimizing positioning of the spacecrafts.

\section{Conclusions}
\label{sec:5}
In this short paper we have shown that we could adjust the orbital characteristics of the three
spacecrafts which consist LISA detector at one order, with respect to $e$, higher than what is 
initially designed, in order to achieve specific kinematical properties. Namely, we could make the breathing mode
of detector arms be optimized with respect to noise induced in the signal through any TDI scheme
used to reduce the noise implications. We have shown, by presenting a few examples, that suitable initial
positioning of the three spacecrafts could reduce the noise, due to lasers, by two orders of magnitude with respect to
initial design. We should note though that since the positioning of spacecrafts  could not be changed throughout
mission's lifetime, only a specific TDI could be highly optimized. If another TDI scheme is used
simultaneously to analyze some signal the benefits of the fine-tuned kinematics will not be equally highlighted.
Hence the choice of kinematics should be based on the TDI scheme that will be most often used in signal analysis.

\begin{acknowledgments}
This research was supported by Grant No 70/4/7672 of the Special Account for Research Grants of the University of Athens.
\end{acknowledgments}


\end{document}